\documentclass[twocolumn,english]{aastex6}
\usepackage[T1]{fontenc}
\usepackage[latin9]{inputenc}
\usepackage{amstext}
\usepackage{amssymb}
\usepackage{graphicx}

\makeatletter

\usepackage{amsmath}
\usepackage{breakurl}
\usepackage{gensymb}

\hypersetup{allcolors=blue}

\journalinfo{The Astrophysical Journal Letters, in press (accepted 2017 December 19)}
\shorttitle{Backtracing Interstellar Objects}
\shortauthors{Zhang}

\makeatother

\usepackage{babel}
\begin{document}

\title{Prospects for Backtracing 1I/\textquoteleft Oumuamua and Future Interstellar
Objects}

\author{Qicheng Zhang}

\email{qicheng@cometary.org}

\affil{Division of Geological and Planetary Sciences, California Institute
of Technology, Pasadena, CA 92115, USA}
\begin{abstract}
1I/\textquoteleft Oumuamua is the first of likely many small bodies
of extrasolar origin to be found in the solar system. These interstellar
objects (ISOs) are hypothesized to have formed in extrasolar planetary
systems prior to being ejected into interstellar space and subsequently
arriving at the solar system. This paper discusses necessary considerations
for tracing ISOs back to their parent stars via trajectory analysis,
and places approximate limits on doing so. Results indicate the capability
to backtrace ISOs beyond the immediate solar neighborhood is presently
constrained by the quality of stellar astrometry, a factor poised
for significant improvement with upcoming \textit{Gaia} data releases.
Nonetheless, prospects for linking 1I or any other ISO to their respective
parent star appear unfavorable on an individual basis due to gravitational
scattering from random stellar encounters which limit traceability
to the past few tens of millions of years. These results, however,
do not preclude the possibility of occasional success, particularly
after considering the potential for observational bias favoring the
discovery of younger ISOs, together with the anticipated rise in the
ISO discovery rate under forthcoming surveys.
\end{abstract}


\section{Introduction}

1I/\textquoteleft Oumuamua (henceforth, 1I) is the first small body
of definitive extrasolar origin to be identified within the solar
system. 1I was originally discovered by the Panoramic Survey Telescope
And Rapid Response System \citep[Pan-STARRS;][]{kaiser2002}, and
announced as C/2017~U1~(PANSTARRS) on 2017~October~25 as a comet
on an improbable trajectory for a solar system object, with eccentricity
$e\approx1.2$ and hyperbolic excess speed $v_{\infty}\approx26$~km~s$^{-1}$
\citep{williams2017a}. The object was subsequently re-designated
A/2017~U1 after follow-up observations revealed a stellar morphology
\citep{williams2017b}, before it was finally given its present designation
on 2017~November~6 under a new dedicated naming scheme for interstellar
objects \citep{williams2017c}.

The anticipated arrival of the Large Synoptic Survey Telescope \citep[LSST;][]{tyson2002},
to be capable of quickly finding fainter objects over wider fields
than current surveys, is poised to dramatically expand the catalog
of similar interstellar objects (ISOs) in the solar system \citep{trilling2017}.
ISOs, whose hypothesized existence far predates the discovery of 1I
\citep{sekanina1976,sen1993,engelhardt2017}, are believed to have
formed in extrasolar planetary systems before being ejected into interstellar
space. Spectral and photometric observations of 1I show an absence
of ultra-red material typical of outer solar system objects, suggesting
it either formed or subsequently evolved within the snow line of its
parent star \citep{ye2017,jewitt2017,meech2017}. In the absence of
a larger sample, the characteristics established for 1I may be assumed
to be representative of typical ISOs.

Given the significant levels ongoing star formation activity in the
Milky Way \citep{kennicutt2012} through which the ISO population
is presumably supplied, it seems conceivable that at least a fraction
of discovered ISOs could potentially be linked back to their parent
stars. Attempts to link 1I to stars in the immediate solar neighborhood
have yielded no obvious candidates \citep{mamajek2017,ye2017}. Additional
searches by \citet{gaidos2017}, \citet{portegies2017}, \citet{dybczynsky2017},
and \citet{feng2017} considered a wider selection of stars, particularly
from the \textit{Tycho}-\textit{Gaia} Astrometric Solution \citep[TGAS;][]{michalik2015},
but mostly focused on the subset of stars with available radial velocity
data for which encounter time and velocity can be constrained. No
definitive point of origin for 1I was identified by any of these searches,
although numerous plausible options have been proposed.

The following sections discuss considerations for backtracing ISOs,
including 1I, to their parent stars in the context of astrometric
uncertainties associated with both the ISO and of the candidate stars.
Approximate bounds are placed on the potential for success in search
attempts for a given ISO, including 1I, in both the present and future.
Finally, a past encounter search method that does not require radial
velocity measurements is introduced and tested for 1I.

\section{Scattering Timescale}

The capacity to predict or, in this instance, backtrace an ISO in
a chaotic dynamical system like the Milky Way galaxy is fundamentally
limited by incomplete information on the state of the system. Specifically,
the close approach of an ISO to a star with gravitational parameter
$\mu_{*}\equiv GM_{*}$ at relative speed $v_{\infty}$ will scatter
the ISO by an angle $\Delta\theta$ given by equation~(\ref{eq:deflect})
that is sensitive to the impact parameter $b$ of the encounter.

\begin{equation}
\sin\frac{\Delta\theta}{2}=\left(1+\left(\frac{bv_{\infty}^{2}}{\mu_{*}}\right)^{2}\right)^{-1/2}\label{eq:deflect}
\end{equation}

Over a typical interstellar distance $L$ between encounters, even
a small $\Delta\theta\ll1$ can produce a large perpendicular displacement
$\Delta y\approx L\Delta\theta\gtrsim b$, resulting in successive
stellar encounters becoming completely unpredictable. In the $\Delta\theta\ll1\implies bv_{\infty}^{2}\gg\mu_{*}$
limit, equation~(\ref{eq:deflect}) simplifies to

\begin{equation}
\Delta\theta=\frac{2\mu_{*}}{bv_{\infty}^{2}}
\end{equation}

Consider the maximum $b=b_{0}$ of a stellar encounter for which the
following encounter of similar $b_{0}$ becomes unpredictable, with
$\Delta y\sim b_{0}$. The average distance between two encounters
of impact parameter below $b_{0}$ in a region of uniform stellar
density $n_{*}$ is given by the corresponding mean free path $L_{0}\sim\left(n_{*}\pi b_{0}^{2}\right)^{-1}$.
In addition, encounter speed $v_{\infty}$ generally satisfies $v_{\infty}\sim u_{*}$,
the stellar velocity dispersion, for a recently introduced ISO. This
approximation is valid as long as the ISO is a galactic disk object,
as 1I presently is. Then,

\begin{equation}
\Delta y=\frac{2\mu_{*}}{\pi b_{0}^{3}n_{*}u_{*}^{2}}
\end{equation}

and $\Delta y\sim b_{0}$ when

\begin{equation}
b_{0}\sim\left(\frac{2\mu_{*}}{\pi n_{*}u_{*}^{2}}\right)^{1/4}
\end{equation}

which corresponds to a mean free path

\begin{equation}
L_{0}\sim\frac{u_{*}}{(2\pi n_{*}\mu_{*})^{1/2}}
\end{equation}

which is crossed over a scattering timescale

\begin{equation}
\tau_{0}\sim\frac{L_{0}}{u_{*}}\sim(2\pi n_{*}\mu_{*})^{-1/2}\label{eq:maxtime}
\end{equation}

This scattering timescale $\tau_{0}$ serves as an approximate upper
bound on the duration over which an ISO can reliably be traced given
near-perfect astrometry of both the ISO and of stars in the galaxy.
Note that to a factor of order unity, the $\tau_{0}$ given by equation~(\ref{eq:maxtime})
also serves as a scattering timescale over which stars, which are
similarly scattered in encounters with other stars, can reliably be
traced.

The solar neighborhood features a wide distribution of $\mu_{*}$
with relative abundance per mass interval $\xi(\mu_{*})$. A characteristic
$\mu_{*}$ occurs at the peak of the $\xi(\mu_{*})\mu_{*}$ distribution,
with stars near this $\mu_{*}$ most strongly constraining $L_{0}$
and $\tau_{0}$ as both values are minimized at maximum $n_{*}\mu_{*}$.
\citet{chabrier2003} finds characteristic values near $\mu_{*}\sim0.2GM_{\odot}$
for stars and close multi-star systems (which are effectively equivalent
to single stars in scattering ISOs) in the Milky Way galaxy.

TGAS, which is nearly complete for stars of this $\mu_{*}$ within
a few parsecs of the Sun, has a stellar density $n_{*}\sim0.15$~pc$^{-3}$
in this region \citep{michalik2015}. In addition, the local stellar
velocity dispersion is estimated to be $u_{*}\sim20$~km~s$^{-1}$
\citep{huang2015}. A trajectory through stellar environments comparable
to the solar neighborhood corresponds to a scattering timescale of
$\tau_{0}\sim30$~Myr, with $L_{0}\sim700$~pc and $b_{0}\sim0.05$~pc.
This result is consistent with that of numerical simulations by \citet{dybczynsky2017}
which find gravitational perturbations from nearby stars to be insignificant.

Suppose ISOs have been produced at a constant rate over the past $t_{\text{max}}\sim10$~Gyr
in the Milky Way galaxy, and are spatially distributed independently
of age. Under this simple model, a fraction $\tau_{0}/t_{\text{max}}\sim3\times10^{-3}$
of these ISOs are expected to be realistically traceable to their
parent stars given near-optimal astrometry together with an accurate
model of the galactic potential.

Note, however, that observational biases favoring young ISOs could
significantly amplify the fraction of traceable ISOs by reducing the
effective $t_{\text{max}}$. Consider an extreme example where a hypothetical
mechanism that removes each ISO from the galaxy after it spends $t_{\text{max}}=10$~Myr
in interstellar space. Since $t_{\text{max}}<\tau_{0}$ for this example,
all discovered ISOs are younger than the scattering timescale, so
have likely never been scattered and remain traceable with accurate
astrometry.

\citet{feng2017} propose a more realistic mechanism by which older
ISOs are scattered out of the galactic disk entirely after repeated
stellar encounters, thus suppressing their discovery rate. Given that
this mechanism requires repeated scattering events, it must operates
over a timescale $t_{\text{max}}\gg\tau_{0}$, and so, still only
a minute fraction of discovered ISOs are likely to be traceable in
the absence of stronger bias.

\section{Astrometric Considerations}

In practice, astrometric uncertainties may limit the range over which
a given ISO can be linked to its parent star well short of the scattering
timescale derived above. Two distinct classes of uncertainties affect
the ability to trace an ISO's trajectory, and both must considered
when attempting to do so:
\begin{enumerate}
\item uncertainties in the ISO's trajectory
\item uncertainties in stellar trajectories
\end{enumerate}
The first is initially large at the ISO's discovery when few observations
are available, then decreases over the course of its passage through
the solar system as additional observations refine its trajectory,
and finally settles at a minimum value once the ISO becomes unobservable.
The second decreases more slowly as improved stellar astrometric catalogs
become available, but may continue to decrease into the foreseeable
future with data from missions like \textit{Gaia}. The consequences
of each class on the potential for backtracing an ISO are discussed
separately below.

\subsection{ISO Trajectory Uncertainties}

Consider, for now, only the uncertainty in the trajectory of the ISO.
In practice, the relevant uncertainty is largely in approach velocity,
with the uncertainty distribution typically taking the form of a flattened
ellipsoid aligned edge-on with the nominal velocity. Let $\Delta\phi_{\text{iso}}$
be the characteristic fractional uncertainty in velocity. If $\Delta\phi_{1}$,
$\Delta\phi_{2}$, and $\Delta\phi_{3}$ are the semi-axes of the
uncertainty ellipsoid, then $\Delta\phi_{\text{iso}}$ is well-represented
by equation~(\ref{eq:isounc})\textemdash chosen such that $\pi\Delta\phi_{\text{iso}}^{2}$
represents a typical cross sectional area for the ellipsoid as approximated
by Thomsen's formula which is generally most accurate with $p\approx1.6075$
\footnote{\url{http://www.numericana.com/answer/ellipsoid.htm}}.

\begin{equation}
\Delta\phi_{\text{iso}}\sim\left(\frac{\Delta\phi_{1}^{p}\Delta\phi_{2}^{p}+\Delta\phi_{2}^{p}\Delta\phi_{3}^{p}+\Delta\phi_{1}^{p}\Delta\phi_{3}^{p}}{3}\right)^{1/2p}\label{eq:isounc}
\end{equation}

The corresponding uncertainty in the ISO's position $\Delta l$ is
related to its distance $r$ from the Sun by $\Delta l\sim r\Delta\phi_{\text{iso}}$.
Given a uniform stellar density $n_{*}$, the mean free path for a
random overlap in position between the ISO an a star is $L\sim(\pi\Delta l^{2}n_{*})^{-1}\sim(\pi r^{2}\Delta\phi_{\text{iso}}^{2}n_{*})^{-1}$.
The nearest random overlap is expected to occur when $r\sim L\to L_{1}$
for which

\begin{equation}
L_{1}\sim(\pi\Delta\phi_{\text{iso}}^{2}n_{*})^{-1/3}\label{eq:idealstdist}
\end{equation}
corresponding to a limiting timescale of

\begin{equation}
\tau_{1}\sim\frac{L_{1}}{u_{*}}\sim(\pi\Delta\phi_{\text{iso}}^{2}n_{*}u_{*}^{3})^{-1/3}
\end{equation}

Beyond $L_{1}$ and $\tau_{1}$, stars passing nowhere near the ISO
are expected to begin crossing the position uncertainty ellipsoid
by chance alone.

\begin{figure}
\includegraphics{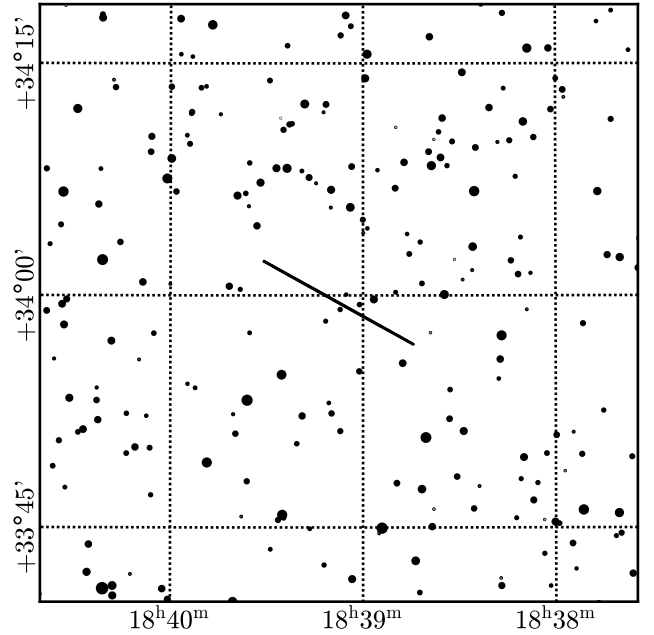}

\caption{Highly elongated 3-sigma uncertainty ellipse (diagonal bar) for the
origin direction of 1I from JPL Small Body Database on 2017 November
20 overlaid on a map of \textit{Gaia} DR1 stars (epoch 2015.0) with
$G$ magnitude~<~14 \citep{gaia2016b}\label{fig:ellipse}}
\end{figure}

As of 2017 November 20, the JPL Small Body Database orbital solution
for 1I\footnote{\url{https://ssd.jpl.nasa.gov/sbdb.cgi?orb=1;sstr=1I}}
indicates a 3-sigma uncertainty ellipse for the origin direction of
1I with semi-axes $\Delta\phi_{1}\times\Delta\phi_{2}=331''\times1.5''$
centered on the J2000.0 coordinates $18^{\text{h}}39^{\text{m}}7^{\text{s}}\llap{.}5$,
$33\degree59'3''\llap{.\,}2$. Figure~\ref{fig:ellipse} plots this
flattened uncertainty ellipse\textemdash a cross section of the full
uncertainty ellipsoid\textemdash over a map of the present-day positions
of bright stars from \textit{Gaia} DR1 for context. The corresponding
uncertainty in radial velocity $v_{\infty}$ is 0.225\% which subtends
an angle $\Delta\phi_{3}=464''$. Equation~(\ref{eq:isounc}) gives
$\Delta\phi_{\text{iso}}\sim280''$ which places limits of $L_{1}\sim100$~pc
and $\tau_{1}\sim5$~Myr for 1I.

Note that for $\Delta\phi_{\text{iso}}\ll\Delta\phi_{\text{sc}}\sim15''$,
$L_{1}\gg L_{0}$ and $\tau_{1}\gg\tau_{0}$. In this scattering-limited
regime, further improvements to the ISO trajectory will not appreciably
extend the range and period over which it can be accurately be tracked
which are ultimately limited by its most recent scattering encounter.
Backtracing is therefore only possible for an ISO whose last scattering
event corresponds to its departure from its origin system.

An ISO ejected from within or near the snow line, as 1I may have been
\citep{ye2017,jewitt2017,meech2017}, would appear to originate in
an encounter with its parent stars at a periastron distance $q\lesssim\delta_{\text{snow}}$,
where $\delta_{\text{snow}}$ is the snow line distance. Consider
$\delta_{\text{snow}}\lesssim10$~au, which holds through the planetesimal
formation process for stars of $\mu_{*}\lesssim3GM_{\odot}$ \citep{kennedy2008}.
Then, for any ISO originating from $r\gg1$~pc, $q$ subtends an
angle of $\phi_{q}\ll10''<\Delta\phi_{\text{sc}}$. At $\Delta\phi_{\text{iso}}\sim\Delta\phi_{\text{sc}}$,
the ISO's departure from its origin system would be indistinguishable
from an exact encounter with the parent star where $q=0$.

The identification and parameters of the last scattering event at
time $t_{\text{ls}}$ before present become well-constrained once
$\tau_{1}\gg t_{\text{ls}}$. Since $t_{\text{ls}}\lesssim\tau_{0}$
if the apparent last scattering is the ISO origin and $t_{\text{ls}}\sim\tau_{0}$
if not, ISO astrometry should aim to reduce uncertainties to $\Delta\phi_{\text{iso}}\ll\Delta\phi_{\text{sc}}\sim15''$
for $\tau_{1}\gg\tau_{0}\sim30$~Myr. Astrometry of such precision
is sufficient to establish the encounter associated with the ISO's
departure as improbable for a random encounter, should this event
be its most recent scattering encounter.

\subsection{Stellar Motion Uncertainties}

The $L_{1}$ and $\tau_{1}$ limits serve as bounds to the long-term
prospects of backtracing a particular ISO after astrometric observations
of the ISO have concluded. In theory, the quality of the stellar astrometry
being searched can continue to improve long after the ISO has left
the solar system, and so uncertainties in this data may become negligible
in the distant future.

However, an actual backtrace done in the present-day is very much
affected by the quality and completeness of the existing stellar astrometry
data. Uncertainties in this data increase the effective cross section
for a possible encounter. Presently, the largest source of stellar
astrometry is the \textit{Gaia} mission \citep{gaia2016a}. Its first
data release, \textit{Gaia} DR1 \citep{gaia2016b} contains 1\,142\,679\,769~stars,
of which, only the 2\,057\,050~stars comprising the TGAS subset
\citep{michalik2015} include a full five-parameter astrometric solution.

The typical 3-sigma proper motion uncertainty among all TGAS stars
is $\epsilon_{\text{pm}}\sim\Delta\phi_{\text{pm}}u_{*}/r\sim4$~mas~yr$^{-1}$,
and the corresponding parallax uncertainty $\epsilon_{\text{plx}}\sim1$~mas.
Let $\alpha\equiv1''$~pc be the proportionality constant relating
parallax and $r_{0}^{-1}$. Then, uncertainty in $r_{0}$ is $\Delta r_{0}\sim r_{0}^{2}\epsilon_{\text{plx}}/\alpha\sim r^{2}\epsilon_{\text{plx}}/\alpha$
which corresponds to a relative uncertainty $\Delta\phi_{r0}\sim r_{0}/r\sim r\epsilon_{\text{plx}}/\alpha$.

Additionally, radial velocity $u_{r}$\textemdash not provided by
TGAS\textemdash is also necessary to determine the timing and relative
speed of a close encounter between an ISO and a star. Let uncertainty
in $u_{r}$ as a fraction of speed be $\Delta\phi_{\text{rv}}$, so
total uncertainty in the radial direction becomes $\Delta\phi_{r}\sim\left(\Delta\phi_{r0}^{2}+\Delta\phi_{\text{rv}}^{2}\right)^{1/2}$.
Then, $\Delta\phi_{r}$ and $\Delta\phi_{\text{pm}}$ are combined
in an analogous fashion to equation~(\ref{eq:isounc}), for an ellipsoid
of $\Delta\phi_{\text{pm}}\times\Delta\phi_{\text{pm}}\times\Delta\phi_{r}$,
into a characteristic stellar uncertainty

\begin{equation}
\Delta\phi_{*}\sim\left(\frac{\Delta\phi_{\text{pm}}^{2p}+2\Delta\phi_{\text{pm}}^{p}\Delta\phi_{r}^{p}}{3}\right)^{1/2p}
\end{equation}
Combining $\Delta\phi_{*}$ with $\Delta\phi_{\text{iso}}$ gives
an effective total relative uncertainty $\Delta\phi\sim\left(\Delta\phi_{\text{\text{iso}}}^{2}+\Delta\phi_{*}^{2}\right)^{1/2}$.
Then, equation~(\ref{eq:idealstdist}) generalizes to

\begin{equation}
L_{2}\sim(\pi\Delta\phi^{2}n_{*})^{-1/3}
\end{equation}

Consider the limit imposed by uncertainties in the five standard astrometric
parameters (i.e., excluding radial velocity), with $\Delta\phi\sim\Delta\phi_{*}$
and $\Delta\phi_{r}\sim\Delta\phi_{r0}$ at $r\sim L_{2}\to L_{2}'$.
Define $\hat{\epsilon}_{\text{pm}}\equiv\epsilon_{\text{pm}}\alpha/u_{*}$,
the proper motion uncertainty in a form that can be related to $\epsilon_{\text{plx}}$.
For TGAS, $\hat{\epsilon}_{\text{pm}}\sim1$~mas~$\sim\epsilon_{\text{plx}}$.
Then,

\begin{equation}
L_{2}'\sim\left(\frac{\alpha^{2}}{\pi n_{*}}\left(\frac{3}{\hat{\epsilon}_{\text{pm}}^{2p}+2\hat{\epsilon}_{\text{pm}}^{p}\epsilon_{r0}^{p}}\right)^{1/p}\right)^{1/5}\label{eq:starlim}
\end{equation}

\begin{figure}
\includegraphics{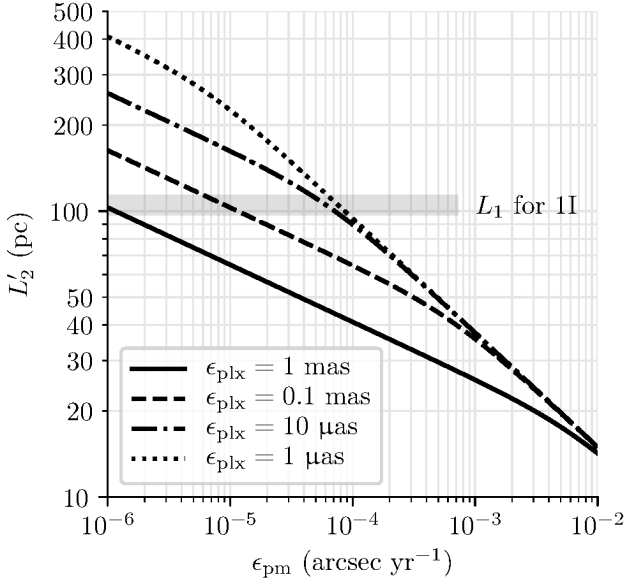}

\caption{Limiting distance $L_{2}'$ for tracing ISOs while constrained by
3-sigma uncertainties in stellar proper motion $\epsilon_{\text{pm}}$
and parallax $\epsilon_{\text{plx}}$, estimated by equation~(\ref{eq:starlim}).
The horizontal bar marks the corresponding $L_{1}$ limit from uncertainties
in 1I's trajectory, as of 2017 November 20, which become constraining
for points above this bar. \label{fig:l2variation}}
\end{figure}

Figure~\ref{fig:l2variation} illustrates the variation in $L_{2}'$
with $\epsilon_{\text{pm}}$ and $\epsilon_{\text{plx}}$, and shows
the potential growth of the traceable region with improved stellar
astrometry in the future. For typical TGAS stars, $L_{2}'\sim20$~pc
, corresponding to $\tau_{2}'\sim L_{2}/u_{*}\sim1$~Myr. However,
stars of $r_{0}<L_{2}'$ are predominantly part of the Hipparcos subset
of TGAS which has a characteristic $\epsilon_{\text{pm}}\sim1$~mas~yr$^{-1}$.
Using this improved $\epsilon_{\text{pm}}$ raises the limits to $L_{2}'\sim30$~pc
and $\tau_{2}'\sim1.5$~Myr. 

ISOs can therefore be reliably traced with TGAS only to encounters
with nearby stars of $r\ll L_{2}'\sim30$~pc. Random stars with no
physical association to the ISO of interest are expected to be matched
beyond this region\textemdash the immediate solar neighborhood\textemdash and
cannot be reliably distinguished from a star actually encountered
by the ISO, let alone its true origin.

\section{Linear Backtracing}

Radial velocity often contribute significantly to the uncertainty
in the state vector of a star, necessary to run a general backtracing
simulation involving nonlinear perturbations. Moreover, these measurements
are available for only a small subset of stars in TGAS. However, while
radial velocity is imperative to constrain the circumstances of an
encounter, its value often has minimal effect on whether an encounter
occurs at all. Five-parameter stellar astrometric solutions are sufficient
to identify encounters with the ISO, provided data uncertainties permit
approximating the motion of the ISO and all considered stars as linear.

Linear motion is a valid approximation over a timescale $\tau_{\text{lin}}$
such that its fraction of the galactic dynamical time, estimated as
the solar orbital period $\tau_{\text{dyn}}\sim200$~Myr \citep{innanen1978},
is comparable to $\Delta\phi$:

\begin{equation}
\frac{\tau_{\text{lin}}}{\tau_{\text{dyn}}}\sim\Delta\phi
\end{equation}

Consider constraints placed by the astrometric uncertainties $\Delta\phi_{\text{pm}}$
and $\Delta\phi_{\text{plx}}$. The condition $r\ll L_{\text{lin}}\sim u_{*}\tau_{\text{lin}}$
is then satisfied as long as $\left(\hat{\epsilon}_{\text{pm}}^{2}+\epsilon_{\text{plx}}^{2}\right)^{1/2}\gg\alpha\left(u_{*}\tau_{\text{dyn}}\right)^{-1}\sim0.2$~mas,
a condition that always holds for TGAS. Approximating motion as linear
therefore should not appreciably increase the total uncertainty, given
present stellar proper motion uncertainties. Note, however, that the
preceding statement may cease to hold for future \textit{Gaia} data
releases should proper motion uncertainties be improved by an order
of magnitude or more as anticipated.

Under the linear motion approximation, two geometric criteria are
needed to isolate the stars with which the ISO may potentially have
had an exact encounter ($b=0$):
\begin{enumerate}
\item The apparent trajectory of every star lies on a great circle. Proper
motion uncertainties expand the single great circle into a distribution
of possible great circle trajectories spanning a pair of wedges, one
leading and one trailing the star's motion. Only those stars where
the trailing wedge sufficiently overlaps the ISO origin uncertainty
ellipse (as shown for 1I in Figure~\ref{fig:ellipse}) can potentially
have had an encounter with the ISO.
\item Let $\tilde{u}$ be the magnitude of a star's proper motion, and $\theta_{0*}$
be the angular separation between the star's current position, and
the point where it crosses the ISO uncertainty ellipse. Only stars
with $r\tilde{u}\equiv u_{\perp}=v_{\infty}\sin\theta_{0*}$, within
data uncertainties, crossed the ISO uncertainty ellipse at the moment
the ISO was at the same distance as the star, and thus correspond
to an encounter in physical space.
\end{enumerate}
The time and distance scales for a random star to satisfy the first
criterion is set by the mean free path to an encounter on the surface
of the celestial sphere. Within a distance $r$, there are $N$ stars:

\begin{equation}
N\sim\frac{4}{3}\pi r^{3}n_{*}
\end{equation}

The angular number density on the celestial sphere is $n_{\Omega}\sim(4\pi)^{-1}N$,
and the corresponding encounter cross section is $\sigma_{\Omega}\sim2\Delta\phi_{\text{pm}}$.
The resulting mean free path is

\begin{equation}
\bar{\theta}\sim(n_{\Omega}\sigma_{\Omega})^{-1}\sim\frac{3}{2r^{3}n_{*}\Delta\phi_{\text{pm}}}
\end{equation}

Random stars begin to cross the uncertainty ellipse at $\bar{\theta}\sim\pi$
which sets the limiting distance $r\sim L_{3\text{a}}$ at

\begin{equation}
L_{3\text{a}}\sim\left(\frac{3}{2\pi n_{*}\Delta\phi}\right)^{1/3}
\end{equation}

Then, with $\Delta\phi_{\text{pm}}\sim\epsilon_{\text{pm}}L_{3\text{a}}/u_{*}$,

\begin{equation}
L_{3\text{a}}\sim\left(\frac{3u_{*}}{2\pi n_{*}\epsilon_{\text{pm}}}\right)^{1/4}\label{eq:crit1dist}
\end{equation}

Using solar neighborhood parameters for $u_{*}$ and $n_{*}$, with
$\epsilon_{\text{pm}}\sim1$~mas~yr$^{-1}$ for the TGAS Hipparcos
subset, gives $L_{3\text{a}}\sim10$~pc, and a corresponding $\tau_{3\text{a}}\sim0.5$~Myr.

The second criterion has the effect of reducing $n_{*}$ by factor
$\kappa_{3\text{b}}$ set by the uncertainty $\epsilon_{\perp}$ in
$u_{\perp}\equiv r\tilde{u}$ relative to the spread $\sigma_{v}$
in $v_{\infty}\sin\theta_{0*}$. Then,

\begin{equation}
\begin{aligned}\epsilon_{\perp} & =u_{\perp}\left(\left(\frac{\epsilon_{\text{pm}}}{\tilde{u}}\right)^{2}+\left(\frac{r\epsilon_{\text{plx}}}{\alpha}\right)^{2}\right)^{1/2}\\
 & \sim\frac{ru_{*}}{\alpha}\left(\hat{\epsilon}_{\text{pm}}^{2}+\epsilon_{\text{plx}}^{2}\right)^{1/2}
\end{aligned}
\end{equation}

Next, with $\sigma_{v}\sim v_{\infty}\sim u_{*}$,

\begin{equation}
\kappa_{3\text{b}}\sim\frac{\epsilon_{\perp}}{\sigma_{v}}\sim\frac{r}{\alpha}\left(\hat{\epsilon}_{\text{pm}}^{2}+\epsilon_{\text{plx}}^{2}\right)^{1/2}
\end{equation}

Finally, substitute $r\to L_{3}$, $n_{*}\to\kappa_{3\text{b}}n_{*}$,
and $L_{3\text{a}}\to L_{3}$ in equation~(\ref{eq:crit1dist}) to
find

\begin{equation}
L_{3}\sim\left(\frac{3\alpha^{2}}{2\pi n_{*}\hat{\epsilon}_{\text{pm}}}\right)^{1/5}\left(\hat{\epsilon}_{\text{pm}}^{2}+\epsilon_{\text{plx}}^{2}\right)^{-1/10}
\end{equation}

which gives $L_{3}\sim30$~pc and $\tau_{3}\sim1.5$~Myr. Since
$L_{3}\sim L_{2}'$ and $\tau_{3}\sim\tau_{2}'$, this method is comparably
capable at identifying encounters as a method using six-parameter
solutions for each star with well-constrained radial velocities. Thus,
as long as uncertainties in stellar astrometry permit the use of the
linear motion approximation, these geometric criteria serve as an
effective means to identify any past encounter of high statistical
confidence with minimal computational power.

\subsection{Search Results}

A search of TGAS for a potential origin of 1I was designed and conducted
with consideration of the uncertainties and limits discussed above.
Stars were filtered by the geometric criteria specified above, with
a successful match is defined by a 3-sigma overlap in uncertainties.
The uncertainty ellipse for the origin direction of 1I is approximated
as completely flattened with minor axis $\Delta\phi_{2}=0$ to simplify
computation. Only stars with nominal distance $r_{0}<r_{\text{max}}=10$~pc
from the Sun\textemdash well within the computed $L_{3}$\textemdash were
considered. None of 68~stars considered matched these criteria, indicating
that the parent star of 1I cannot be conclusively identified with
TGAS.

This null result does not necessarily imply that 1I originated from
beyond this region. While TGAS contains most stars in the immediate
solar neighborhood of the characteristic $\mu_{*}\sim0.2GM_{\odot}$
and above, it does not constitute a comprehensive catalog of planetary
systems in this region. Even brown dwarfs\textemdash completely absent
in TGAS\textemdash have been observed with protoplanetary disks from
which ISOs could plausibly originate \citep{apai2005}, and moreover,
may be as abundant as stars \citep{chabrier2003}. The full \textit{Gaia}
DR1 samples these objects, with full astrometric solutions expected
in upcoming data releases, albeit with limited completeness beyond
a few parsecs \citep{debruijin2014}.

Note that including these low mass stars and substellar brown dwarfs
raises $n_{*}$\textemdash perhaps doubling its value\textemdash but
does not appreciably change the estimated $L_{1}$, $L_{2}'$, and
$L_{3}$ (and their corresponding timescales) which vary extremely
slowly with $n_{*}$. These objects should also contribute only minimally
to ISO scattering and $L_{0}$ which, as discussed earlier, is primarily
constrained by stars of $\mu_{*}\sim0.2GM_{\odot}$.

\begin{figure}
\includegraphics{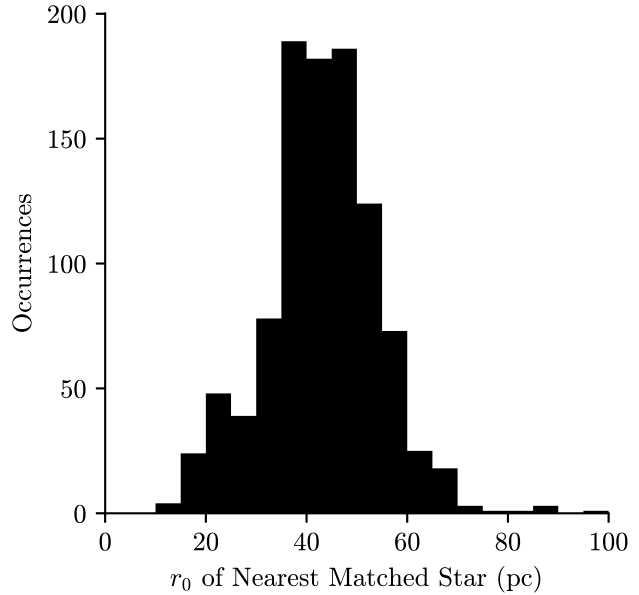}

\caption{Distribution of nominal distance $r_{0}$ for the nearest star matched
by the introduced geometric search criteria (bin size: 5~pc) over
a sample of 1000 pseudo-randomly generated ISOs with approach velocities
distributed according to the local stellar velocity dispersion, but
with relative uncertainties ($\Delta\phi_{1}$, $\Delta\phi_{2}$,
and $\Delta\phi_{3}$) matching those of 1I's trajectory. This distribution
is consistent with the predicted $r_{0}\sim L_{3}\sim30$~pc for
the nearest match in a typical search. \label{fig:searchtgas}}
\end{figure}

Finally, as a consistency check of the estimated limiting distance
$L_{3}$, the search was repeated with all stars in TGAS (i.e., $r_{\text{max}}\to\infty$).
The nearest star matching both geometric criteria\textemdash presumably
by random chance\textemdash had a nominal $r_{0}\approx40.5$~pc.
This distance is fairly close to the $r_{0}\sim L_{3}\sim30$~pc
expected of the nearest random matching star for an arbitrary ISO
with trajectory uncertainties $\Delta\phi_{1}$, $\Delta\phi_{2}$,
and $\Delta\phi_{3}$ identical to those of 1I.

To ensure that this result is not unique to 1I's trajectory, this
procedure was repeated, replacing 1I with a sample of 1000~ISOs with
pseudo-randomly generated nominal approach trajectories. These trajectories
were produced by selecting its galactic $U$, $V$, and $W$ components
of velocity from normal distributions matching those of the local
stellar velocity dispersion \citep{huang2015}. Results are plotted
in Figure~\ref{fig:searchtgas} which shows that the distribution
of $r_{0}$ for the nearest star matching the criteria is approximately
normal, with mean $42.9\pm0.4$~pc and standard deviation $11.2\pm0.3$~pc.
These results are, again, roughly consistent with the analytically
estimated $r_{0}\sim L_{3}\sim30$~pc for the nearest match of a
typical search, to within a factor of two.

\section{Conclusions}

1I is the first, but likely not the last ISO to be discovered in the
solar system. Forthcoming surveys like LSST will likely provide additional
opportunities to study these objects and, in effect, the distant environments
in which they form. The capability to match an ISO to its original
planetary system would provide a unique mode to examine specific extrasolar
planetary systems at close proximity.

The preceding sections showed that trajectory analysis alone is unlikely
to be successful for any particular ISO, although the possibility
of occasional success cannot be excluded. Gravitational scattering
from random stellar encounters limits traceability to the past few
10~Myr. Observational bias favoring young ISOs may elevate the fraction
of discovered ISOs that are potentially traceable above the <1\% of
galactic history contained by this limit. More stringent limits, however,
are placed by astrometric uncertainties associated with both the ISO
and the stars. These uncertainties become important when velocity
uncertainties greatly exceed one part in $10^{5}$\textemdash an angular
uncertainty of a few arcseconds\textemdash including in the present
case of 1I.

Currently, stellar astrometry provides the limiting constraint, with
TGAS enabling the positive identification of a close encounter out
to just a few parsecs, spanning only the immediate solar neighborhood.
A search conducted for 1I of TGAS stars presently within this region
fails to produce such a positive identification, indicating that the
ISO must have originated from either a nearby low mass star or brown
dwarf not included in TGAS, or from a more distant star outside the
region that cannot, at present, be identified. This result does not
preclude searches for candidates beyond the limited region\textemdash including
those found by \citet{gaidos2017}, \citet{portegies2017}, \citet{dybczynsky2017},
and \citet{feng2017}\textemdash but merely implies that such candidates
are unlikely to be linkable to the ISO's trajectory with reasonable
statistical confidence. Improvements to stellar astrometric uncertainty
are anticipated with the upcoming \textit{Gaia} data releases which
are expected to significantly expand this range and improve catalog
completeness in the near future.

\acknowledgements
Thanks to Quan-Zhi Ye, Shreyas Vissapragada, Yayaati Chachan, and
Konstantin Batygin for insightful discussions on the potential origins
of 1I. Special thanks to an anonymous reviewer whose comments and
suggestions helped improve this manuscript. This research has made
use of data and/or services provided by the International Astronomical
Union's Minor Planet Center, and by the Jet Propulsion Laboratory's
Solar System Dynamics Group. This work has made use of data from the
European Space Agency (ESA) mission {\it Gaia} (\url{https://www.cosmos.esa.int/gaia}),
processed by the {\it Gaia} Data Processing and Analysis Consortium
(DPAC, \url{https://www.cosmos.esa.int/web/gaia/dpac/consortium}).
Funding for the DPAC has been provided by national institutions, in
particular the institutions participating in the {\it Gaia} Multilateral
Agreement.

\software{Matplotlib \citep{hunter2007}, NumPy \citep{walt2011}, Python \citep{vanrossum1995}}

\end{document}